\documentclass[12pt]{article}
\usepackage{amsmath}
\usepackage{amssymb}
\begin{document}

\begin{center}

{\bf \noindent  VISCOUS DARK COSMOLOGY WITH ACCOUNT OF QUANTUM
EFFECTS}

\vspace{1cm} I. BREVIK\footnote{E-mail: iver.h.brevik@ntnu.no} and
O. GORBUNOVA\footnote{On leave from Tomsk State Pedagogical
University, Tomsk, Russia. Email: gorbunovaog@yandex.ru.}

\bigskip

 \noindent Department of Energy and Process Engineering, Norwegian
University of Science and Technology,
 N-7491 Trondheim, Norway

 \bigskip


 \end{center}

\begin{abstract}
The analytic properties of the energy density $\rho(t)$ of the
cosmic fluid, and the Hubble parameter $H(t)$, are investigated
near to the future singularity $t=t_s$ assuming different forms
for the equation of state. First, it is shown that the inclusion
of quantum effects coming from the conformal anomaly modifies the
singularity. Thereafter, we consider the effect coming from a bulk
viscosity in the fluid. The viscosity tends to reduce the
magnitude of $t_s$, but does not alter the singularity itself (the
exponent). Main emphasis is laid on the simple case when the
equation of state is $p=w\rho$, with $w$ a constant.
\end{abstract}


\section{Introduction}

It is actually not more than about five years since the possible
occurrence of a singularity of the universe in the far future -
usually called the Big Rip - was first discussed in the literature
\cite{caldwell03,mcinnes02,barrow04}. To begin with, two different
types of singularities were investigated. Later, it became clear
that there are two possibilities more, so that present we know
about four different types. For convenience, we quote from
Ref.~\cite{nojiri05} the following classification:

(i) Type I ("Big Rip"): For $t \rightarrow t_s,~ a\rightarrow
\infty,~\rho \rightarrow \infty$, and $|p|\rightarrow \infty$, or
$p$ and $\rho$ are finite at  $t=t_s$.

(ii) Type II ("sudden"): For $t\rightarrow t_s,~a\rightarrow
a_s,~\rho \rightarrow \rho_s$, and $|p|\rightarrow \infty$,

(iii) Type III: For $t\rightarrow t_s,~a\rightarrow a_s,~\rho
\rightarrow \infty$, and $|p|\rightarrow \infty$,

(iv) Type IV: For $t\rightarrow t_s,~a\rightarrow a_s,~\rho
\rightarrow 0,~|p|\rightarrow 0$, or $p$ and $\rho$ are finite.
Higher order derivatives of $H$ diverge.

Here the notation is standard, $a$ meaning the scale factor and
$t_s$ referring to the instant of the singularity. These theories
assume the cosmic fluid to be ideal, i.e. nonviscous.

As often happens in physics, when the mathematical formalism
encounters singularities, this indicates that the description has
been too simplistic. It becomes natural to ask: can the
singularity be modified or softened by drawing other physical
effects into consideration? The answer turns out to be
affirmative, due to at least two reasons. The first is the effect
of quantum corrections due to  the {\it conformal anomaly}. It
turn up in various cosmological applications: dynamical Casimir
effect with conformal anomaly \cite{brevik99}, or dark fluid with
conformal anomaly \cite{nojiri03}. Explicit calculation shows this
anomaly to cause a softening of the singularity
\cite{nojiri05,nojiri04}. A second reason - the one to be treated
in the present paper - is to take into account the {\it bulk
viscosity} of the cosmic fluid. From a hydromechanical standpoint
a generalization of the cosmic theory so as to encompass viscosity
is most natural. As we will see, the viscosity also acts in the
direction of softening. Viscous cosmology, beginning with the
seminal paper of Misner \cite{misner68} seems generally to have
attracted increased interest in recent years. Some references in
this direction are
\cite{padmanabhan87,gron90,brevik04,ren06,capozziello06,brevik06,sussman08}.

For convenience, and for reference purposes, we give in the next
section a brief overview of the singularity application of
conformal anomaly theory. Thereafter, from Section~3 onwards we
embark on the core of the present paper, namely to include a bulk
viscosity in the Einstein equations.

 \section{Quantum effects, when  $p=-\rho-A\rho^{\beta}$.}

We shall content ourselves by considering specific examples. Close
to the borderline between the quintessence and phantom regions,
the equation-of-state  (EOS) parameter $w=p/\rho$ is near to  -1.
One natural choice for the equation of state is the following
\cite{nojiri04a,nojiri05,nojiri05a,nojiri07}:
\begin{equation}
p=-\rho-A\rho^\beta, \label{1}
\end{equation}
where $A$ and $\beta$ are constants. Thus $p$ always diverges when
$\rho$ becomes infinite. Note that this may be considered as a
viscous dark energy.

We shall distinguish between two cases, viz.  $1/2<\beta \le 1$,
and $\beta >1$. They may be classified as follows, referring to
the notation in the previous section:

(1)  $1/2<\beta \le 1:$ The singularity is of Type I. The dominant
energy condition (DEC), requiring $\rho \ge 0, ¨\rho \pm p \ge 0$,
is broken. The EOS parameter approaches -1 for either sign of $A$,
when $\rho \rightarrow \infty$. This case implies that $\rho$
diverges in a finite future. One obtains \cite{nojiri05}
\begin{equation}
\rho \sim (t_s-t)^{\frac{2}{1-2\beta}}, \quad H \sim
(t_s-t)^{\frac{1}{1-2\beta}}, \label{2}
\end{equation}
with $t_s$ the singularity time and $H=\dot{a}/a$ the Hubble
parameter.

(2)  $\beta >1:$ The singularity is of Type III. The DEC is also
now broken. The magnitude of the EOS parameter goes to infinity,
i.e., $w\rightarrow +\infty ~(-\infty)$ for $A<0 ~(A>0)$.
Equations (\ref{2}) still hold.

We shall be concerned mainly with the case $\beta >1$ in this
section.

\bigskip

Consider next the quantum contribution to the conformal anomaly.
The complete energy density is $\rho_{tot}=\rho+\rho_A$. Taking
the trace of the conformal anomaly energy-momentum tensor,
$T_A=-\rho_A+3p_A$, plus observing the  energy conservation law,
\begin{equation}
\dot{\rho}_A+3H(\rho_A+p_A)=0, \label{3}
\end{equation}
we find that
\begin{equation}
p_A=-4\rho_A-\dot{\rho}_A/H. \label{4}
\end{equation}
Thus we obtain for the conformal anomaly energy density
\cite{nojiri05}
\[ \rho_A=-\frac{1}{a^4}\int dt a^4HT_A \]
\[=-\frac{1}{a^4}\int
dta^{4}H[-12b\dot{H}^{2}+24b'(-\dot{H}^2+H^{2}\dot{H}+H^{4})\]
\begin{equation}
-(4b+6b'')(\dddot{H}+7H\ddot{H}+4\dot{H}^{2}+12H^{2}\dot{H})],
\label{5}
 \end{equation}
 where $b, b'$ and $b''$ are constants, occurring in the
 expression for the conformal trace anomaly,
 \begin{equation}
 T_A=b(F+\frac{2}{3}\Box R)+b'G+b''R. \label{6}
 \end{equation}
 Here $F$ is the squared Weyl tensor and $G$ the Gauss-Bonnet
 invariant. Explicitly, if there are $N$ scalars, $N_{1/2}$
 spinors, $N_1$ vectors, $N_2$ gravitons, and $N_{HD}$ higher
 derivative conformal scalars, one has for $b$ and $b'$ the
 expressions
\[ b=\frac{N+6N_{1/2}+12N_{1}+611N_{2}-8N_{HD}}{120(4\pi)^{2}},
 \]
 \begin{equation}
b'=\frac{N+11N_{1/2}+62N_{1}+1411N_{2}-28N_{HD}}{360(4\pi)^{2}},\label{7}
\end{equation}
whereas $b''$ is an arbitrary constant whose value depends on the
regularization.

The quantum corrected FRW equation is, with $\kappa^2=8\pi G$,

\begin{equation} \frac{3}{\kappa^{2}}H^{2}=\rho+\rho_{A} \label{8}.
 \end{equation}
As $R$ is large near $t=t_{s}$, we assume
$(3/\kappa^2)H^{2}\ll|\rho_{A}|.$  Hence, $\rho\sim -\rho_{A}$.

In the presence of quantum effects, suppose that
\begin{equation}
\rho=\rho_{0}(t_{s}-t)^{\tilde{\gamma}}\label{9}
 \end{equation}
as in Ref.~\cite{nojiri05}.  From the energy conservation equation
\[
H=\frac{\dot{\rho}}{3A(\rho)^\beta}\simeq-\frac{\tilde{\gamma}\rho_{0}^{1-\beta}}{3A}(t_{s}-t)^{-1+\tilde{\gamma}(1-\beta)}\]
we get, by taking the most singular term, $\dot{\rho}\sim -(4b+
6b'')H \dddot{H}$.  Integrating with respect to $t$ we get $\rho$
to be of the form (\ref{9}), with $\tilde{\gamma}=4/(1-2\beta).$
Thus
\begin{equation}
 \rho\sim(t_{s}-t)^{\frac{4}{1-2\beta}}, \quad
H\sim(t_{s}-t)^{\frac{3-2\beta}{1-2\beta}} \label{10}
\end{equation}
near $t=t_{s}$. If we compare this with the case of no quantum
corrections, we see that the energy density diverges more rapidly
in the present case, whereas   $H$ becomes less singular. That
means,  the quantum corrections moderate the singularity. Note
that
\begin{equation}
\frac{H^2}{\rho} \sim (t_s-t)^{\frac{2-4\beta}{1-2\beta}}
\rightarrow 0, \label{11}
\end{equation}
showing that our approximation neglecting the left hand side of
Eq.~(\ref{8}) is consistent.

 The case  $\beta >1$ as assumed here, corresponds to singularity of Type III.
  The same properties may be found  for the Big Rip (Type I)
singularity.

\section{Viscosity-generated singularity, when $p=-\rho -f(\rho)$}

Assume now that the cosmic fluid is viscous, with a bulk viscosity
$\zeta$. In general, $\zeta=\zeta(\rho)$. Hereafter  we leave out
the conformal anomaly, so that   $\rho_A=0$. The FRW equation
becomes accordingly $(3/\kappa^2)H^2=\rho$, where the matter
energy density $\rho$ incorporates the modification due to the
viscosity. We assume the cosmic fluid to be thermodynamically a
simple, one-component, fluid. We start by assuming the general
form $p=-\rho-f(\rho)$, with $f(\rho)$ a general function.

This case was analyzed in Ref.~\cite{brevik05}. The essential
difference from the foregoing is that the energy conservation
equation is now
\begin{equation}
\dot{\rho}+3H(\rho+p)=9\zeta H^2 \label{12}
\end{equation}
(note that the scalar expansion is $\theta=3H$). We let $t=0$
denote present time, and let subscript 0 refer to present time
quantities. The viscosity-generated singularity time is denoted as
$t_{s\zeta}$. From Eq.~(9) in \cite{brevik05} we have
\begin{equation}
t=\frac{1}{\sqrt{3}\,\kappa}\int_{\rho_0}^\rho
\frac{d\rho}{\sqrt{\rho}f(\rho)[ 1+\kappa
\zeta(\rho)\sqrt{3\rho}/f(\rho) ]}. \label{13}
\end{equation}
Singularities of Type I or III occur if $t=t_{s\zeta}$ corresponds
to $\rho \rightarrow \infty$.

We assume henceforth that $\zeta(\rho) \rightarrow \zeta$ is a
constant, and consider two specific choices for the function
$f(\rho)$.

\section{The case when $f(\rho)=\alpha \rho$}

We consider this case first because it is simplest. It corresponds
to an equation of state in the form
\begin{equation}
p=w\rho=-(1+\alpha)\rho, \label{14}
\end{equation}
with $\alpha$ a constant. From Eq.~(17) in \cite{brevik05} we have
\begin{equation}
H(t)=\frac{H_0e^{t/t_c}}{1-\frac{3\alpha}{2}H_0t_c(e^{t/t_c}-1)},
\label{15}
\end{equation}
where $t_c$ is a 'viscosity time',
\begin{equation}
t_c=(\frac{3}{2}\kappa^2\zeta)^{-1}. \label{16}
\end{equation}
It is convenient to introduce two new nondimensional quantities
$x$ and $K$,
\begin{equation}
x=\frac{t_{s\zeta}}{t_c}, \quad K=\frac{2}{3\alpha H_0
t_{s\zeta}}. \label{17}
\end{equation}
From Eq.~(\ref{15}) it is seen that $H$ becomes infinite when
\begin{equation}
x=\ln (1+Kx). \label{18}
\end{equation}
This transcendental equation for $x$ can have two solutions.

{\it (i) Nonviscous solution}. There is always the solution $x=0$,
corresponding to $t_c=\infty$ or $\zeta=0$. This is accordingly
the  solution for a nonviscous fluid. We will call the singularity
time in this case $t_{s0}$. We get
\begin{equation}
t_{s0}=\frac{2}{3\alpha H_0}. \label{19}
\end{equation}
It is desirable to write the solutions such that the difference
from the singularity time occurs explicitly. First, we have
\begin{equation}
H(t)=\frac{H_0 t_{s0}}{t_{s0}-t}, \label{20}
\end{equation}
which corresponds to
\begin{equation}
a(t)=\frac{a_0 \,t_{s0}^{2/3\alpha}}{(t_{s0}-t)^{2/3\alpha}}.
\label{21}
\end{equation}
Next, as $H^2 \propto \rho$ according to the FRW equation, we get
\begin{equation}
\rho(t)=\frac{\rho_0\,t_{s0}^2}{(t_{s0}-t)^2}. \label{22}
\end{equation}
Comparing Eqs.~(\ref{20}) and (\ref{22}) with Eq.~(\ref{2}), we
see that corresponding equations for $H(t)$ and $\rho(t)$ are of
the same form, if $\beta=1$. This is as we might expect, as the
equation of state in the present case corresponds to $A=\alpha,\,
\beta=1$. The singularity is of Type I if $\alpha >0$, and of type
III if $\alpha <0$.

{\it (ii) Viscous solution}. In the viscous case, $x>1$,
Eq.~(\ref{18}) has a second root if the slope $K$ is larger than
one. As $K$ is actually the same as the ratio $t_{s0}/t_{s\zeta}$,
we thus get the following condition for solution
\begin{equation}
t_{s\zeta}<t_{s0}. \label{23}
\end{equation}
The presence of a bulk viscosity thus {\it reduces} the future
singularity time. It would not be so easy to foresee this property
beforehand.

For a given value of $t_c$, the transcendental equation (\ref{18})
does not allow us to solve for $t_{s\zeta}$ explicitly. However,
we may easily handle analytically the limiting case $t\rightarrow
t_{s\zeta}$, which is the situation of main interest here. Using
the approximation $ e^{\tau/t_c}=1+\tau/t_c$ with
$\tau=t_{s\zeta}-t$, we obtain
\begin{equation}
H(t)\rightarrow \frac{H_0\,t_{s0}}{t_{s\zeta}-t}, \quad
t\rightarrow t_{s\zeta}. \label{24}
\end{equation}
This formula holds for all values of $\zeta$.

It is thus apparent that we get the same singular behavior of
$H(t)$ in the viscous case as we did previously, in
Eqs.¨(\ref{20}) and (\ref{2}) (the latter under the assumption
that $\beta=1$). Correspondingly, we get in the present case
\begin{equation}
a(t) \sim (t_{s\zeta}-t)^{-2/3\alpha}, \quad t\rightarrow
t_{s\zeta}, \label{25}
\end{equation}
\begin{equation}
\rho(t) \sim (t_{s\zeta}-t)^{-2}, \quad t\rightarrow t_{s\zeta}.
\label{26}
\end{equation}
Whereas the viscosity tends to shorten the future singularity
time, it does {\it not} modify the exponents in the singularity.
Again, the singularity is of Type I if $\alpha >0$, and of type
III if $\alpha <0$.

When dealing with viscosity we have introduced two different
times, namely a viscosity time $t_c$ defined in Eq.~(\ref{16}),
and a singularity time $t_{s\zeta}$. One may wonder: can anything
be said about the relative magnitude of these times? As most
cosmological theories are formulated without viscosity, it is
natural to assume that the cases of main physical interest pertain
to setting $\zeta$ very small, or $t_c$ very large. Accordingly,
the quantity $x$ defined in Eq.~(\ref{17}) should be expected to
be small, corresponding to $t_{s\zeta} \ll t_c$.

Another related point is to inquire whether there is a
relationship between the singularity time  $t_s$ with or without
conformal anomaly considered in Section 2, and the viscosity
singularity time $t_{s\zeta}$.  Although there does not seem to be
a close relationship, it is in our opinion most natural to
identify the nonviscous  time $t_{s0}$  in Eq.~(\ref{19}) with the
previous time $t_s$. As we have seen, the effect from bulk
viscosity is to
 reduce the singularity time. We thus suggest that the
 relationship
 \begin{equation}
 t_{s\zeta} <t_s \label{27}
 \end{equation}
 is quite general.

 Note that one may take into account quantum effects in this case
 too. However, the corresponding equations are very cumbersome and
 require numerical study. Such a study indicates that that quantum
 effects may again soften the future singularity.

 \section{Remarks on the  case when $f(\rho)=A\rho^\beta$ }

 In this more general case we go back to Eq.~(\ref{13}) (still
 assuming that $\zeta$ is constant). Leaving at first the integration limits
  unspecified, we have
 \begin{equation}
 t= \frac{1}{3\kappa^2}\frac{1}{\zeta}\int
 \frac{d\rho}{\rho \left[ 1+\frac{A}{\sqrt{3}\,\kappa \zeta}\,
 \rho^{\beta-1/2}\right]}. \label{28}
 \end{equation}
As $\beta >1/2$, it is seen that the last term in the
 denominator (the viscosity term) dominates for large $\rho$. We
 perform the integration from an initial value of $\rho$ already
 lying in this region, up to $\rho \rightarrow \infty$. The
 corresponding time integration limits are $t$ and $t_{s\zeta}$.
We obtain approximatively
\begin{equation}
t_{s\zeta}-t=\frac{1}{\sqrt{3}\,\kappa}\,\frac{1}{A}\,\frac{1}{\beta-1/2}\,\frac{1}{\rho^{\beta-1/2}}.
\label{29}
\end{equation}
Thus
\begin{equation}
\rho(t)\sim (t_{s\zeta}-t)^{\frac{2}{1-2\beta}}, \quad
t\rightarrow t_{s\zeta}, \label{30}
\end{equation}
which generalizes Eq.~(\ref{26}) and reduces to it in the case
when $\beta=1$. Equation (\ref{30}) has actually the same form as
Eq.~(\ref{2}).

\section{Summary}

We may summarize our results as follows.
 \vspace{0.5cm}

1) We started out by giving a brief overview of the four types of
future singularities, listing in Eq.~(\ref{2}) the known forms for
$\rho(t)$ and $H(t)$  in the case where the equation of state is
assumed as in Eq.~(\ref{1}). Actual in our context were
singularities of Type I and Type III.

2) Inclusion of quantum effects coming from the conformal anomaly
- cf. Eq.~(\ref{5}) - caused the exponents of the future
singularity  to be modified.

3) In Section 3 we left out quantum effects, but considered
instead the influence from a bulk viscosity $\zeta$. Assuming the
simple form (\ref{14}) for the equation of state, we showed in
Eqs.~(\ref{24})-(\ref{26}) that the viscosity acts so as to
shorten the singularity time, but it does {\it not} change the
nature of the singularity. The same property was found when
dealing with a more complicated equation of state in Section 5.
The explicit calculations were limited to the case when $\zeta$
was a constant.

Finally we mention that, under quite general conditions,
preliminary numerical work on quantum effects indicates that they
tend to soften the future singularity.

\section*{Acknowledgment}

The work of O. G. was supported in part by RFBR grant 06-01-00609
and LRSS project N.2553.2008.2.


\end{document}